\documentclass[aps,prl,twocolumn,preprintnumbers,amsmath,amssymb,superscriptaddress,floatfix]{revtex4-1}
\usepackage{graphicx}% Include figure files
\usepackage{dcolumn}% Align table columns on decimal point
\usepackage{bm}% bold math

\usepackage[T1]{fontenc}
\usepackage[english]{babel}
\usepackage{amsmath}
\usepackage{amssymb}
\usepackage{amstext}
\usepackage{bbold}
\usepackage{braket}
\usepackage{relsize}
\usepackage{subfigure}
\usepackage{hhline}
\usepackage{booktabs}
\begin{document}

\title{Entanglement entropy for the long range Ising  chain}
\author{Thomas Koffel}\email{thomas.koffel@icfo.es}
\affiliation{ICFO -- Institut de Ci\`{e}ncies Fot\`{o}niques, Parc Mediterrani de la Tecnologia, 08860 Castelldefels, Spain}

\author{M. Lewenstein}\email{maciej.lewenstein@icfo.es}
\affiliation{ICFO -- Institut de Ci\`{e}ncies Fot\`{o}niques, Parc Mediterrani de la Tecnologia, 08860 Castelldefels, Spain}
\affiliation{ICREA-Instituci\'{o} Catalana de Recerca i Estudis Avancats, 08010 Barcelona, Spain}

\author{Luca Tagliacozzo}\email{luca.tagliacozzo@icfo.es}
\affiliation{ICFO -- Institut de Ci\`{e}ncies Fot\`{o}niques, Parc Mediterrani de la Tecnologia, 08860 Castelldefels, Spain}

\date{\today}

\begin{abstract}
We consider the Ising model in a transverse field with long-range antiferromagnetic  interactions that decay as a power law with their distance. We study both the phase diagram and the entanglement properties as a function of the exponent of the interaction. The phase diagram can be used as a guide for future  experiments with trapped ions.  We find two gapped phases, one dominated by the transverse field, exhibiting quasi long range order, and one dominated by the long range interaction, with long range N\'eel ordered ground states. We determine the location of the  quantum critical points separating those two phases. We determine their critical exponents and central-charges. In the phase with quasi long range order the ground states exhibit exotic corrections to the area law for the entanglement entropy coexisting with  gapped entanglement spectra.
\end{abstract}
\maketitle

%\section{Introduction}

Long range (LR) interactions have attracted a lot of attention since they could produce interesting new phenomena\cite{ruelle_statistical_1968,dyson_existence_1969,cardy_one-dimensional_1981,lahaye_physics_2009} . Recently there have been impressive advances in controlling experimentally quantum systems. In particular it has been shown 
by Britton {\it et al.} \cite{britton_engineered_2012} that beryllium ions can be stored in a Penning trap, where an accurate laser design can induce  LR Ising anti-ferromagnetic interactions among  them. This is only the most recent of a series of impressive experimental results on using trapped ions to simulate spin models \cite{friedenauer_simulating_2008,kim_quantum_2010,islam_onset_2011}. Motivated by these results we analyze the phase diagram of the anti-ferromagnetic LR  Ising Hamiltonian in the presence of a transverse field (LITF). The difference with the standard Ising model in a transverse field (ITF) is that the two-body part of the Hamiltonian includes interactions among arbitrary separated pairs of spins, whose strength decays as a power law of their distance $r$, $r^{-\alpha}$ with $\alpha >0$.

For the LITF we  i) the determine  the full phase diagram of the model as a function of $\alpha$, that can be used as a guide for future experiments with trapped ions ii) quantify the increase of complexity induced by the LR interaction for the classical simulation of the model  and iii) characterize the phase transitions.

Regarding both i) and iii), we identify two different phases. One of them, dominated by the local part  of the Hamiltonian,  is gapped and presents patterns of  quasi-long range order (QLRO) induced by the LR part of the Hamiltonian. This is exotic, since normally QLRO is associated to gapless phases. The other, dominated by the LR terms of the Hamiltonian, presents anti-ferromagnetic LR order (LRO) in the form of  N\'eel ground states. Between them, we observe a line of quantum phase transitions, whose nature depends on the value of $\alpha$. They either  are  in the same universality class than the ITF ($\alpha>2.25$), or present new universal behaviors (for $\alpha\le 2.25$). 

Concerning ii),  we focus on the entanglement entropy content of the ground states of the LITF  as a function of both  $\alpha$ and the size of the system. A common belief, (see, however, Ref. \onlinecite{evenbly_real_2012} for an updated perspective)  relates the amount of entanglement contained in a state with its simulability with classical computers \cite{vidal_efficient_2004, tagliacozzo_simulation_2009}. This practically translates into the fact that  those states that obey the ``area law'' for the entanglement,   can be simulated classically, since  their entanglement  scales only with the area  of a region  rather than with its volume. In particular  all  ground states of gapped short range (SR) Hamiltonians in 1D obey the ``area law'' \cite{hastings_area_2007, masanes_area_2009}. For ground state of  LR Hamiltonians one can expect a different scenario. Indeed, we show that in some cases their ground state still obey the area law. More interestingly, in the phase with QLRO, we observe unusual  violations to it, in a gapped phase, where the local part of the Hamiltonian is dominant.

Our studies complement the existing one in several  ways. On one side most of the quantum many body literature has focused on  LR dipolar interactions decaying with the distance as  $r^{-3}$ \phantom{a}  \cite{porras_effective_2004,deng_effective_2005,hauke_complete_2010,peter_anomalous_2012,nebendahl_improved_2012,wall_out_2012}. Much less has been done for a generic LR interaction of the type $r^{-\alpha}$ as the one we consider here \cite{cannas_long-range_1996,dutta_phase_2001,dalmonte_2010}. For these systems, even less has been done with respect to the interplay between anti-ferromagnetism and LR interactions. This is particularly interesting since  the anti-ferromagnetic  ITF is equivalent through rotation of one every two spins  to the ferromagnetic ITF, while this is not the case for the LITF. In few cases the effect of LR interactions has been considered on top of a LRO N\'eel state, but the interaction  considered was  non-frustrating with respect to the LRO \cite{laflorencie_critical_2005,sandvik_ground_2010}. In the case we consider here, all the frustration comes from the anti-ferromagnetic nature of the LR interaction.
 
As a side result, we have improved current Matrix Product states (MPS) techniques. Indeed we have generalized the time dependent variational  principle (TDVP) so that it can be used with LR interactions (alternative approaches can be found in the literature \cite{deng_effective_2005,mcculloch_infinite_2008, crosswhite_applying_2008,hauke_complete_2010,nebendahl_improved_2012,wall_out_2012}). The generalization is described in detail in the appendix.  While the choice of using an MPS ansatz for LR systems could be questioned (since MPS are best suited for ground states of local Hamiltonians \cite{verstraete_matrix_2006}),  our results validate this choice (see also the  work \cite{latorre_entanglement_2005}). Indeed we observe that  the strongest violations to the area law are  logarithmic in the system size as  for SR critical points where ground states can still be represented efficiently with MPSs \cite{verstraete_matrix_2006}. 

Qualitatively, however, the logarithmic corrections seem to coexist with a gapped entanglement spectrum, a very exotic feature. Indeed our data suggest that the ES could present bands and gaps, even if we cannot exclude the possibility of just one gap separating the first eigenvalue from a continuum of them. 

% The structure of the paper is organized as following, in section \ref{sect:model} we introduce the model and the methods used to study it. Section \ref{sect:res} contains most of the  numerical results. They are divided in three main subsections,  we first discuss the various phases by considering the scaling of the entanglement entropy and characterizing the ES in each of them. We then concentrate on the location of  the  phase transition as a function of $\alpha$, and extract various quantities of theoretical interest for its classification such as the central charge, and some of the critical exponents.

\emph{The model.}
\label{sect:model}
We study   a one dimensional spin chain with open boundary conditions (OBC). We analyze the ground state of the system described by the  LITF  Hamiltonian 
\begin{equation}
 H(\theta, \alpha)=  \sin (\theta) \sum_{i,j} \frac{1}{|i-j|^{\alpha}} \sigma_x^i \sigma_x^j + \cos( \theta) \sum_{i} \sigma_z^i \label{eq:ham},
\end{equation}
where $i, j$ are two arbitrary points of the 1D chain, $\alpha \ge 0$. We consider the  anti-ferromagnetic  phase, $0\le \theta \le \frac{\pi}{2}$.
The reasons for that i) it  is the interesting regime for the experimental results in  \cite{britton_engineered_2012},  ii) we are interested in studying the interplay between LRO and frustration; iii) the ferromagnetic LITF has been already  studied  elsewhere \cite{dutta_phase_2001,ruelle_statistical_1968,dyson_existence_1969}. 

%For values of  $\alpha$ in the range $1\ < \alpha \le \frac{5}{3}$ the  ferromagnetic LITF has two phases, the phase transition separating them has mean field exponents given by $\gamma =1, \nu = \frac{1}{\alpha-1}, \eta=3 -\alpha$. For $ \alpha < \frac{1}{2}$ the model has just one phase where the ground state is given by a GHZ state in the basis where $\sigma_x$ is diagonal. For $\alpha > 2.25$ the ground state is essentially the same than the one of  ITF model. In between the mean-field regime and the SR regime  there is a region of $\alpha$ and the  phase transitions separating them  present  exotic critical exponents \cite{dutta_phase_2001}. 

The frustration effects \cite{hauke_complete_2010} prevent us from using  standard Quantum Monte Carlo  so that we turn to   matrix product state (MPS)  techniques \cite{mcculloch_density-matrix_2007}. We  use a variational algorithm (known as TDVP \cite{haegeman_time-dependent_2011}) to obtain numerically the best possible MPS for the ground state of \ref{eq:ham}. In order to deal with  the LR,  the Hamiltonian is encoded  in a matrix product operator (MPO) \cite{mcculloch_density-matrix_2007}. This requires an extension of the original TDVP algorithm (described in the appendix). Alternative techniques are also available \cite{mcculloch_infinite_2008,crosswhite_applying_2008, nebendahl_improved_2012}. 

 In order to establish the phase diagram and  to locate the phase transitions we study the behavior of the entanglement entropy defined as 
\begin{equation}
  S_{L/2} =  - \textrm{tr} \rho_{L/2} \log \rho_{L/2}, \label{eq:ent}
\end{equation}
where $\rho_{L/2} = \textrm{tr}_{i_1 \cdots i_{L/2}} \ket{\Omega}\bra{\Omega}$ and $\ket{\Omega}$ is the ground state of the system. The Hamiltonian \ref{eq:ham} has a ${\cal Z}_2$ symmetry generated by ${\cal G}=\prod_i \sigma_z^i$.  For $\theta >\theta_c$ the two body terms of the Hamiltonian dominate. They only commute with $G$ globally so  that the spectrum of the reduced density matrix $\rho_{L/2}$ (neglecting spontaneous symmetry breaking effects) is doubly degenerate \cite{perez-garcia_matrix_2007}. For $\theta < \theta_c$ on the other hand the local part of  \ref{eq:ham} dominates. It commutes locally with $G$ so that the spectrum  becomes non-degenerate. Close to the  change of degeneracy we observe a maximum of $S_{L/2}$ that we use as the signature for the phase transition.

We then analyze the entanglement spectrum (ES) on both sides of the transition. It is defined in terms of the logarithm of the reduced density matrix
\begin{equation}
 h_i = \log(\rho_i),\label{eq:ent_spect}
\end{equation}
where $\rho_i$ are the eigenvalues of $\rho_{L/2}$.
 For $\theta > \theta_c$  the ES  can be fully described by using perturbation theory (PT). For $\theta < \theta_c$ we observe both a perturbative and a non-perturbative regime for the ES depending on the value of  $\alpha$. In the non-perturbative regime  we observe in it the  appearance of bands.  In the same phase the entanglement entropy violates the area law by  exhibiting scaling with respect to the system size.

Once we identify the critical point we consider the finite size scaling of the correlation functions 
\begin{equation}
\langle \sigma_x^{L/2}  \sigma_x^{L/2+L/5}\rangle  \propto L^{-2 \Delta_{x}}, \  \langle \sigma_z^{L/2}  \sigma_z^{L/2+L/5}\rangle  \propto L^{-2 \Delta_{z}}. \label{eq:corr}
\end{equation}
The corresponding exponents, as a function of $\alpha$ present two different regimes. A SR regime, where the critical exponents are the ones of the ITF, and a LR regime, where the exponents vary continuously with $\alpha$.

\emph{Numerical results} 
\label{sect:res} 
We have performed several TDVP simulations of finite chains with length $L$ in the range $20<L<150$ and OBC. The  interactions encoded in the MPOs correctly reproduce the desired  power law $r^{-\alpha} $ in the range  of distances  $1\le r \le L/2$ \footnote{As discussed in detail in \cite{mcculloch_density-matrix_2007,crosswhite_applying_2008,frowis_tensor_2010}, using matrix product operators allows to easily encode exponentially decaying LR interactions. Power law decays are obtained approximately by expanding the interaction onto a series of exponentials.}. For each simulation, we have increased  the MPS bond dimension $\chi$ up to convergence of the ground state energy to ten digits when passing from one value of $\chi$ to the next one $\chi'$ (typically $\chi'=2 \chi$). This typically happens at values of $\chi \le 100$.

\begin{figure}
\includegraphics[width=7cm]{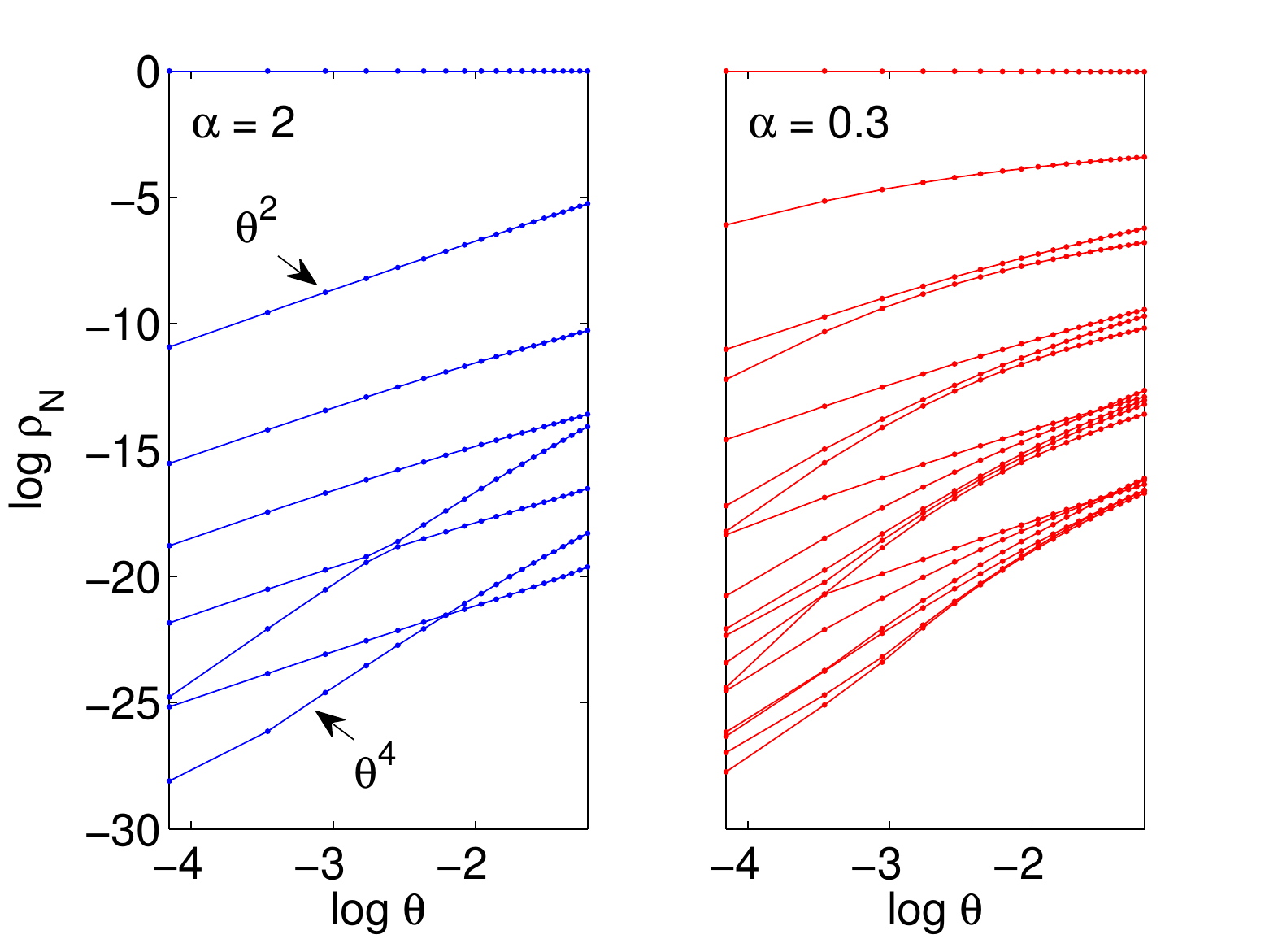}
 \caption{\textbf{Structure of the ES}  defined in Eq. \ref{eq:ent_spect}  as a function of $\log (\theta)$,  for $0 \le \theta \le  0.36$, deep in the $z$ polarized phase\label{fig:ent-spect}. \textbf{Left panel} $\alpha=2$, the spectrum presents well separated scales reproducible by the lowest order in PT. Several eigenvalues belong to the same order in PT ( that can be distinguished as parallel lines ) contrary to what happens in the ITF. \textbf{Right panel},  the ``non-perturbative'' regime for $\alpha=0.3$ ,  there is no clear scale separation, and no clear power law dependence of the eigenvalues with $\theta$. The eigenvalues tend to cluster to form bands  separated by gaps.}
\end{figure}

\emph{Phase diagram.}
In the anti-ferromagnetic case for all values of $\alpha >0$ the system shows two phases. For values of $\theta \sim 0$, $\theta \ll \theta_c(\alpha)$, the ground state $\ket{\Omega}$  can be understood as a perturbative  modification of the  product state locally pointing along the $-1$ eigenvector of $\sigma_z$. In formulas, defining $\sigma_z \ket{\uparrow}=\ket{\uparrow}$, $\sigma_z \ket{\downarrow} = -\ket{\downarrow}$, 
$
 \ket{\Omega}_{\theta =0} =\prod_{i} \ket{\downarrow}^i,
$
is independent of the value of $\alpha$. This is a gapped phase, where elementary excitations are spin-flips. For values of   $\alpha \le 1$ and $\theta \ll \theta_c(\alpha)$,  the ground state starts to encode patterns of correlations induced by the LR part of the Hamiltonian that suggest the existence of a non-perturbative regime (see  Fig \ref{fig:ent-spect} right panel). However, when passing from the perturbative regime to the non-perturbative regime, none of the observables we have considered shows an  anomalous behavior, so that we conclude  that there is no sharp phase transition  between them.

For all the values of $\alpha$  we have considered, at some $\theta_c(\alpha)$ the system undergoes a second order phase transition to a predominantly N\'eel ordered state aligned in the $x$ direction. At the fixed point of the N\'eel  phase (at $\theta=\frac{\pi}{2}$), the ground of the system is independent of $\alpha$,   
$
 \ket{\Omega}_{\theta =\frac{\pi}{2}} = \frac{1}{\sqrt{2}}\left(  [\ket{+}\ket{-}]  \cdots + [\ket{-}\ket{+}]  \cdots \right) ,
$
where $\sigma_x \ket{+}=\ket{+}$ and $\sigma_x \ket{-}=-\ket{-}$, and the square brackets indicate the elementary two-sites unit cell.
% that repeats up to the total system size (that we suppose a multiple of the basic unit-cell).
The first excited states in these phases are kinks. The gap to them vanishes as $\alpha$ approaches zero.  At $\alpha=0$, indeed, the 1D geometry of the system is completely lost and the N\'eel state  melts into an exponentially degenerate ground-state subspace made of all possible arrangements of $N/2$ $\ket{+}$ states and $N/2$ $\ket{-}$ . 

The value of  $\theta_c(\alpha)$ is always larger than the one of the ITF transition at $\pi/4$.
This can be understood intuitively:  the  slower the two body interaction decays, (smaller $\alpha$) the more the  $\sigma_x$ part of the Hamiltonian becomes frustrated. As a consequence, for small values of the transverse field ( large  values of $\theta$) the $z$ polarized state has lower energy than the highly frustrated  N\'eel state so that the system transitions to  $z$ polarized phase. This  also explains why  in the $(\theta, \alpha)$  $\theta_c $ increases with decreasing $\alpha$. 

\begin{figure}
 \includegraphics[width=7cm]{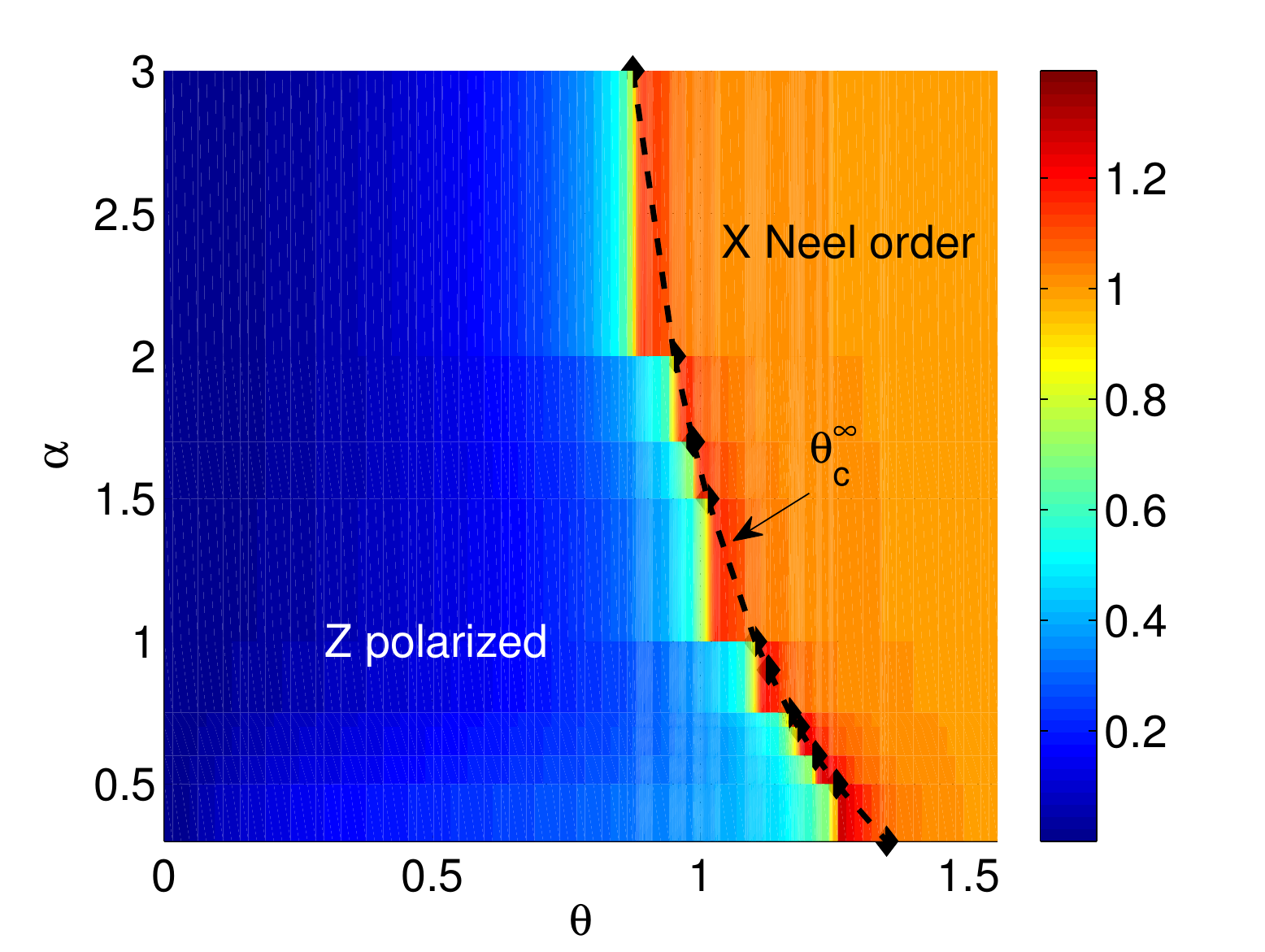}
\caption{\textbf{ Phase diagram of the LITF from the entanglement entropy.} The half chain entanglement entropy provides information about the phase diagram of  \ref{eq:ham} as a function of $\theta$ and   $\alpha$.  The background  colors represent the value $S_{L/2}$ for a system of size $L=100$ and several values of $\alpha$ in the range $0.3\le \alpha \le 3$ . The maximum of it, signals the vicinity of a phase transition. Extrapolating its position as a function of $L$, for $L=20 \cdots 100$, we locate the position of the transition in the thermodynamic limit $\theta_c^{\infty}$. The results are  superimposed  as  solid black dots connected by a dashed line.  The anti-ferromagnetic nature of the LR interaction favors the $z$ polarized phase so that the transition always occurs on the right of $\theta=\pi/4$, the critical value for the ITF.\label{fig:phase_dia} }
\end{figure}

The phase diagram is presented in Fig. \ref{fig:phase_dia}, where we plot $S_{L/2}$ of Eq. \ref{eq:ent} as a function of both $\alpha$ and $\theta$. For fixed $L$ and $\alpha$,   $S_{L/2}$ has a maximum at some given $\theta^*$. By extrapolating the values of $\theta^*$ as a function of $L$ we determine the location of the critical point  $\theta_c^{\infty}(\alpha)$. These points  are superimposed to the colored background data for $S_{L/2}$ at $L=100$ in black in Fig. \ref{fig:phase_dia} and are joined by a dashed line as a guide to the eye.

\begin{figure}
\includegraphics[width=7cm]{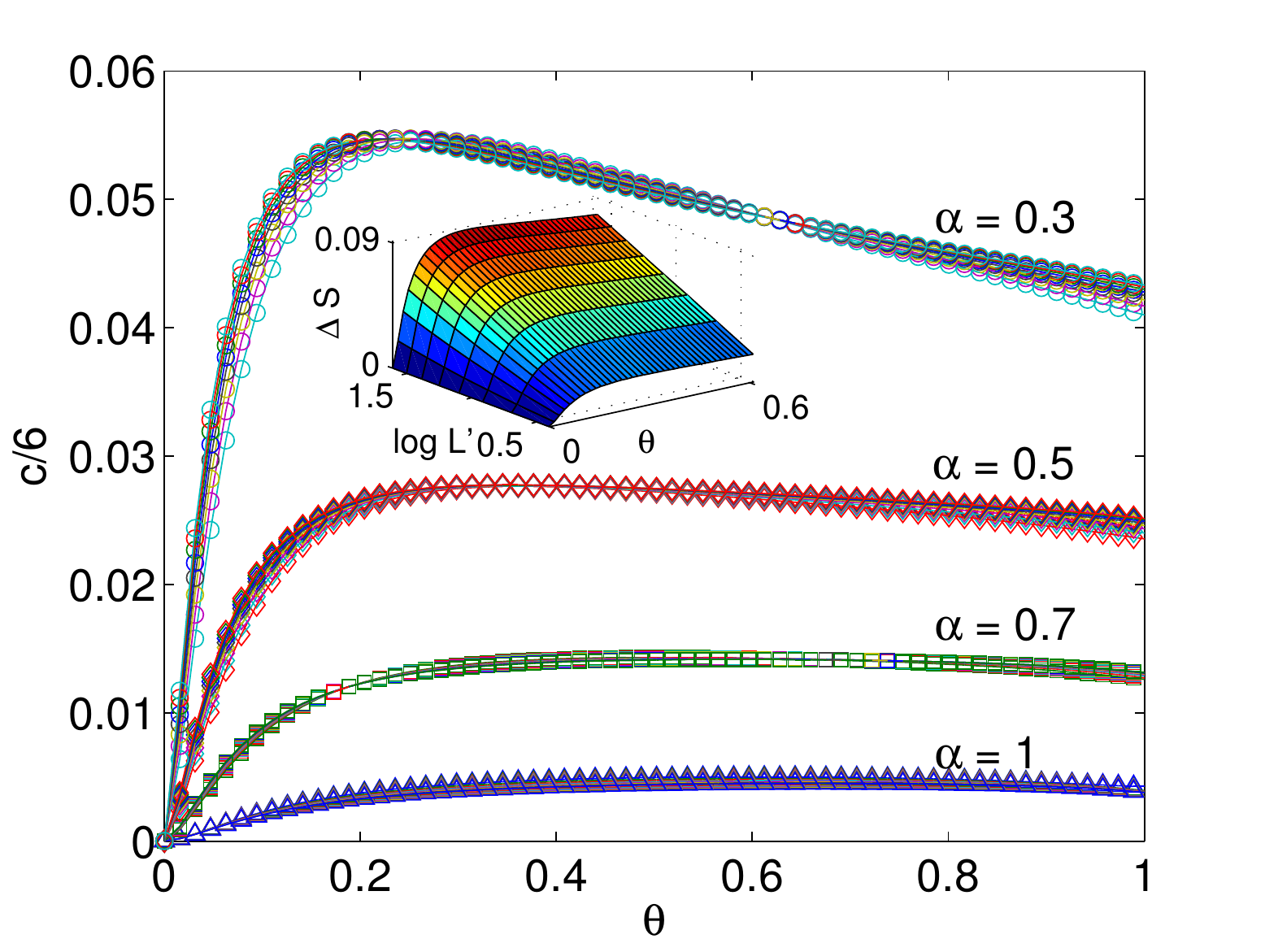}
 \includegraphics[width=7cm]{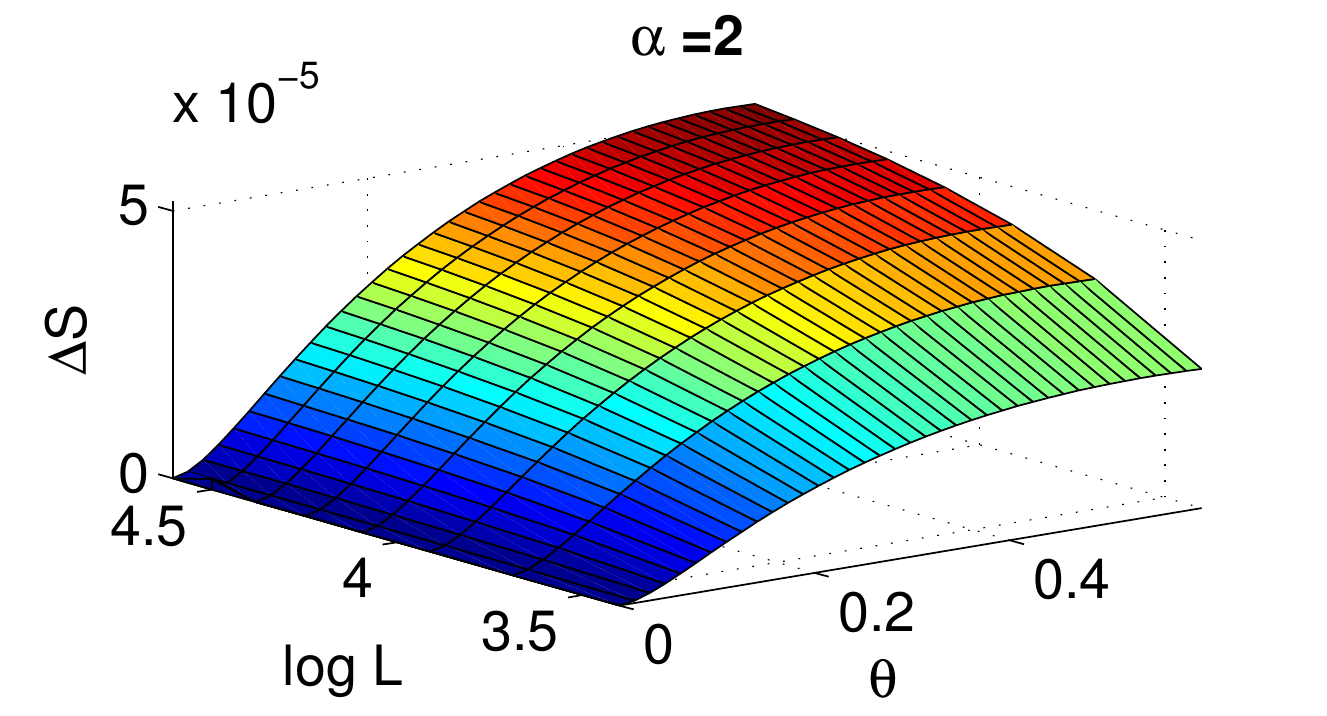}
\caption{\textbf{Violations to the area law.} The entanglement entropy for a bipartition increases monotonically with the system size in the whole $z$ polarized phase. \textbf{Upper panel.} For $\alpha \le 1$ the entropy scales logarithmically with the size of the system (small inset). The pre-factor  in Eq. \ref{eq:ent-scaling} is extracted by plotting   $c/6=\frac{\Delta S}{log L'}$, with $\Delta S = S_{L/2}-S_{L_0/2}$ and $L'=L/L_0, L_0=20, L=30\cdots 100$. \textbf{Lower panel.} For $\alpha=2$ the scaling is sub-logarithmic (but still monotonically increasing).\label{fig:area_law}} 
\end{figure}

In the $z$ polarized phase, we observe two very striking phenomena. On one side, even if the phase is gapped,  we observe polynomially decaying correlation functions. Namely $\langle \sigma_x^{L/2} \sigma_x^{L/2+r}\rangle \propto r^{-\alpha}$ while $\langle \sigma_z^{L/2} \sigma_z^{L/2+r}\rangle \propto r^{-2 \alpha}$ (similar results were also obtained  in  \cite{hauke_complete_2010,peter_anomalous_2012}). On the other side we observe violations to the area law for the entanglement entropy,
%In one dimension the area law implies that for large enough block the entropy should saturate to a constant value independent on the size of the block. On the contrary 
since of the entropy increases without saturation with the size of the blocks we have considered. There are two different regimes for the violations depending on the value of  $\alpha$. For $\alpha\le 1$ we observe logarithmic violations to the area law so that, by using a tempting analogy with the case of critical systems, \cite{calabrese_entanglement_2004,callan_geometric_1994,latorre_ground_2003} we can define an ``effective central charge'' as 
\begin{equation}
 S_{L/2} \propto \frac{c}{6} \log L.\label{eq:ent-scaling} 
\end{equation}
The value we determine for  $c/6$ are reported in the upper panel of Fig. \ref{fig:area_law}. They are extracted  by  plotting  $\Delta S = S_{L/2}-S_{L_0/2}$ divided by  $L'=L/L_0$, with $L_0=20$ (we use as a reference size   to eliminate the  possible constant terms in the scaling). The small dispersion of the curves obtained from different system sizes  $L$ around a single curve is a confirmation   of the correctness  of the scaling form \ref{eq:ent-scaling}.   Interestingly, this effective central-charge, in the non-perturbative regime, varies very slowly with  $\theta$.
For  $\alpha > 1$ we still observe a steady growth of the entanglement with the size of the blocks but its behavior is sub-logarithmic as shown in the lower panel of Fig. \ref{fig:area_law}. Our data for  $\alpha=3$ are not conclusive.  They  suggest also there the presence of  sub-logarithmic corrections for the sizes considered but they are so slow that  we  cannot exclude that the entropy would eventually saturate for larger systems. We leave this as an open issue. 

The ES of Eq. \ref{eq:ent_spect} can be used to distinguish between  the perturbative and the ``non-perturbative'' regime in the $z$ polarized phase.
 In the perturbative regime, (for $\theta\simeq 0$),   the ES shows well defined scale separation, proportional to different powers of the small parameter $\theta$. The elements of the spectrum are  dominated by the leading order at which they appear in the calculation. In the   ITF  there is a single element at each order in PT, 
%This means  that  $h_1$ is close to 1, and then the main contribution to $h_2$ is at order  $\theta^2$, the one for $h_3$ at order  $\theta^6$, and so on.
whereas in the  LITF  ES instead,  multiple eigenvalues appear at the same order in PT. They can be identified as parallel straight lines by  plotting the ES in a log-log plot as a function of $\theta$. The  slopes of them  indicate to which order in PT  the eigenvalue belongs to, as shown in Fig. \ref{fig:ent-spect} right panel for  $\alpha=2$. There  we appreciate  both the proliferation of eigenvalues and the wide range of  validity of PT. We also see that the ES  is dominated by eigenvalues appearing  at most at order $\theta^4$ in PT. 

In the same range of $\theta$, the ES for $\alpha=0.3$  looks very different. In Fig. \ref{fig:ent-spect} right panel, we  do not see neither a clear separations of scales, nor a well-defined power-law behavior of the eigenvalues with respect to $\theta$ both footprints of the ``non-perturbative'' regime \footnote{We would like to stress that we do not exclude that one could extract the ES by  higher order computation in PT with respect to $\theta$  but rather than the behavior is very different from the one of the perturbative regime.}.  The eigenvalues tend to cluster in bands (and the respective gaps) that are robust to changes in size (at least for the range of sizes we have access to).  An  unsolved issue is whether they would survive to the thermodynamic limit.

  In both  perturbative and  non-perturbative regime the logarithmic violations to the area law coexist with a gapped ES. This gap is likely to survive in the thermodynamic limit,  so that these  corrections are different from those of a quantum critical point, where the ES gap closes with the system size \cite{calabrese_entanglement_2008} .

\emph{The phase transition.}
For the anti-ferromagnetic interaction we find a phase transition for every value of $\alpha >0$. This has to be compared with the ferromagnetic case where there is a lower critical dimension $\alpha=1/2$ \cite{dutta_phase_2001}.  A mean field analysis  around the ITF critical point \cite{fisher_critical_1972,cardy_one-dimensional_1981,cardy_scaling_1996} suggests that the LR interactions are relevant for $\alpha < 2 + 2 \Delta_x^{SR}$ driving the system to a different critical point than the SR case,  with $\Delta^{SR}_x =1/8$ being the scaling dimension of the $\sigma_x$ operator for the SR ITF.  For $\alpha = 2.25$ the LR is marginal, while for $\alpha>2.25$ it  becomes irrelevant and one should observe the standard SR ITF criticality. 

\begin{figure}
 \includegraphics[width=7cm]{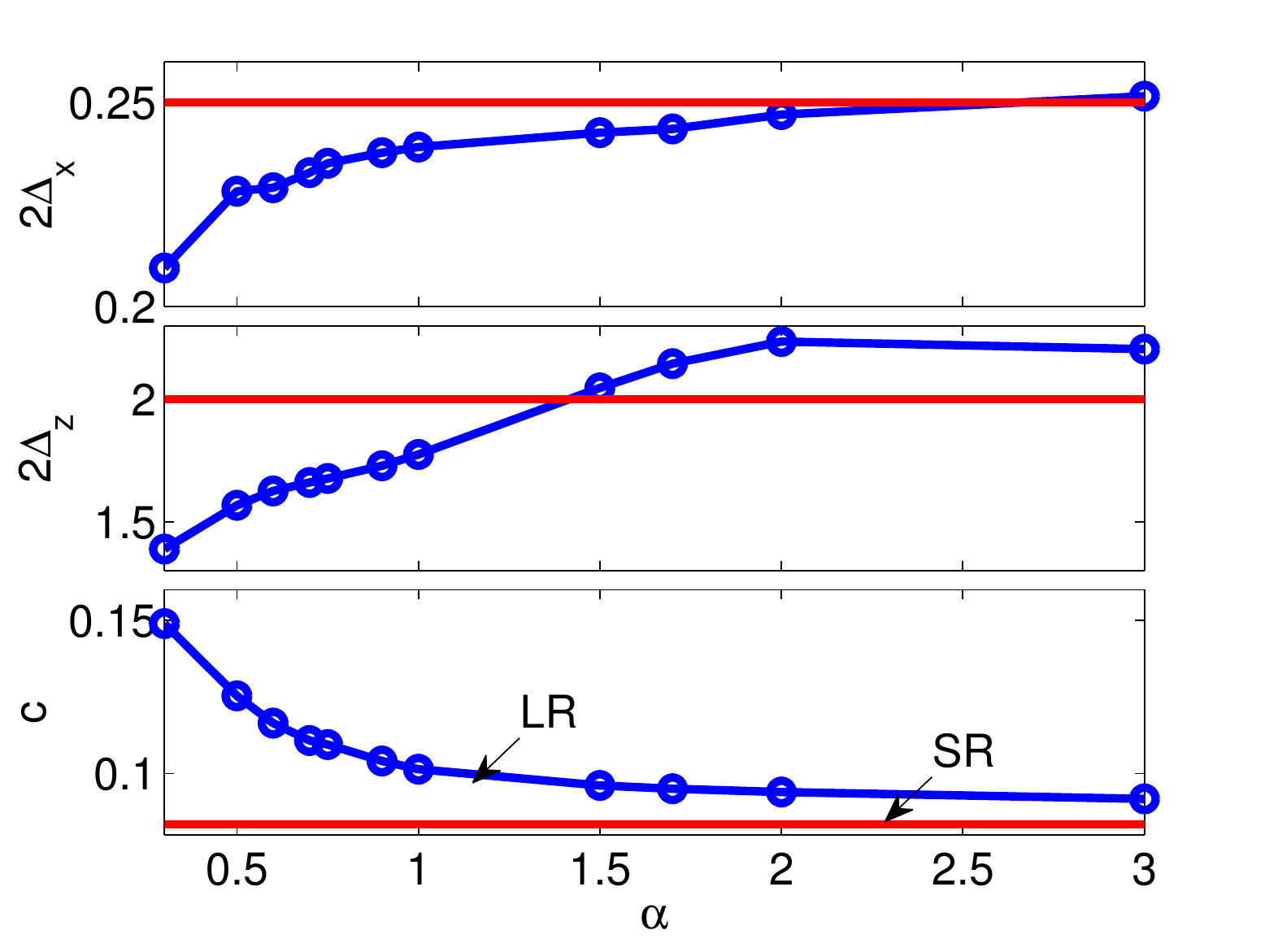}
\caption{ \textbf{Long range universality class.} From a mean field analysis  for  $\alpha>2.25$ the LR  become irrelevant.  \textbf{Upper panel},   $2 \Delta_x ^{LR}$ as a function of $\alpha$. As expected close to $\alpha= 2$ the exponent tends to its SR value $1/4$ . \textbf{Middle panel}, $2 \Delta_z ^{LR}$ does not coincide with the expected thermal exponent $2$ due to subleading corrections.   \textbf{Lower panel}. The central charge extracted from  $S_{L/2}$,  unexpectedly,  is systematically larger than $1/2$. \label{fig:crit_exp} }
\end{figure}

We check  the above statements  performing a finite-size scaling analysis  of the correlation functions \ref{eq:corr},  $ \braket{\sigma_x^{L/2}, \sigma_x^{L/2+L/5}}\propto L^{-2 \Delta_{x}^{LR}}$ . The exponents $\Delta_x^{LR}(\alpha)$  are  presented in the upper panel of Fig.  \ref{fig:crit_exp}, $2 \Delta_x^{LR}(\alpha)$ is different from $2 \Delta^{SR}$ for all values of $\alpha<2$  while between $2$ and $3$ it becomes very close to expected SR value $1/4$.  

By studying the scaling of $S_{L/2}$ in Eq. \ref{eq:ent-scaling}, we can extract the value of the central charge of the corresponding CFT that,  for the ITF, is $c=1/2$. In the whole range of $\alpha$ considered, the coefficient we obtain is systematically bigger than $1/2$. The reason for that is not clear but probably resides in a mixture of effects,  i) the  effects of boundaries  are enhanced by the LR interaction,  ii) the system sizes we can address  are still too small to get rid of the irrelevant contributions to the leading scaling \cite{cardy_unusual_2010} (indeed our data agree with a pure logarithmic scaling only for the biggest lattices $L=70 \cdots 100$) , iii) the LR could induce some marginal operator inducing corrections to the scaling difficult to control\cite{cardy_unusual_2010}. The corresponding plot is presented in the lower panel of Fig. \ref{fig:crit_exp} .

 Finally we have checked the leading power-law scaling of the $\sigma_z$ correlation,  an operator that already for the ITF is not a scaling field on its own. There we expect that its leading scaling is dictated by the thermal exponent  $\Delta_z^{SR}=1$. The results for the LITF are presented in  the central panel of  Fig. \ref{fig:crit_exp}. In the SR regime, for $\alpha >2.25$, the exponent we extract from the fit gives an estimate of the thermal exponent   off by around  $10\%$ clear symptom of contamination with sub-leading corrections.

% \begin{figure}
%  \caption{Location of the critical point for the Hamiltonian of \ref{eq:ham} as a function of $\alpha$ \label{fig:crit}.}
% \end{figure}

\emph{Conclusions and Outlook.}
In this paper we have considered the effects of a LR anti-ferromagnetic interaction on the phase diagram of the ITF, in order to both  provide a guide to future  trapped ions experiments and study the increase of complexity induced by the LR interactions. The resulting phase diagram shows that  the frustration favors the $z$ polarized phase over the $x$ aligned N\'eel phase. For all values of $\alpha>0$ considered we have located  the  phase transition. There we have confirmed that the LR interaction  is relevant for $\alpha \le 2.25$, inducing critical exponents different from the ones of the ITF.   We have determined them for the $\sigma_x \sigma_x$ and $\sigma_z \sigma_z$ correlations (the equivalent of the magnetic and thermal exponents in the SR case). They vary continuously as a function of $\alpha$ in the range $0<\alpha<2.25$. 
The scaling of the entanglement entropy  in the SR regime is used to provide  an estimate for the central charge $c$ of the underlying CFT, that turns out to be systematically larger than the expected value $1/2$. We miss a complete understanding of this result (that however could be a manifestation of the fact that the system we can address are still too small to see the expected asymptotic scaling) and further studies should be devoted to clarify it. 

It is worth to mention that the complexity of the ground state induced by the LR part of the Hamiltonian is not significantly higher than the one of their SR equivalent. However we have encountered surprising violations to the area law for the entanglement entropy, whose strength depends on $\alpha$. The strongest violations  are found for  $\alpha\le 1$ and are  logarithmic in the system size. These violations that only appear in the SR dominated $z$ polarized gapped phase, seem to be always accompanied by a finite entanglement gap and in some cases by the presence of bands in the ES. Further studies should be devoted to check  the persistence of the corrections for dipolar interactions in the thermodynamic limit.  In the  $x$ aligned N\'eel phase, dominated by the  LR part of the Hamiltonian, there are no violations to the area law. LT acknowledges early discussions with F. Cucchietti and P. Hauke on the topic, and financial support from the  Marie Curie project FP7-PEOPLE-2010-IIF ENGAGES 273524 . We also acknowledge the correspondence with J. Haegeman and P. Calabrese.  

\bibliographystyle{apsrev4-1}
%\nocite{*}
\bibliography{long-range}

\appendix
\section{Time dependent variational principle with long range interactions}
\begin{figure}
 \includegraphics[width=7cm]{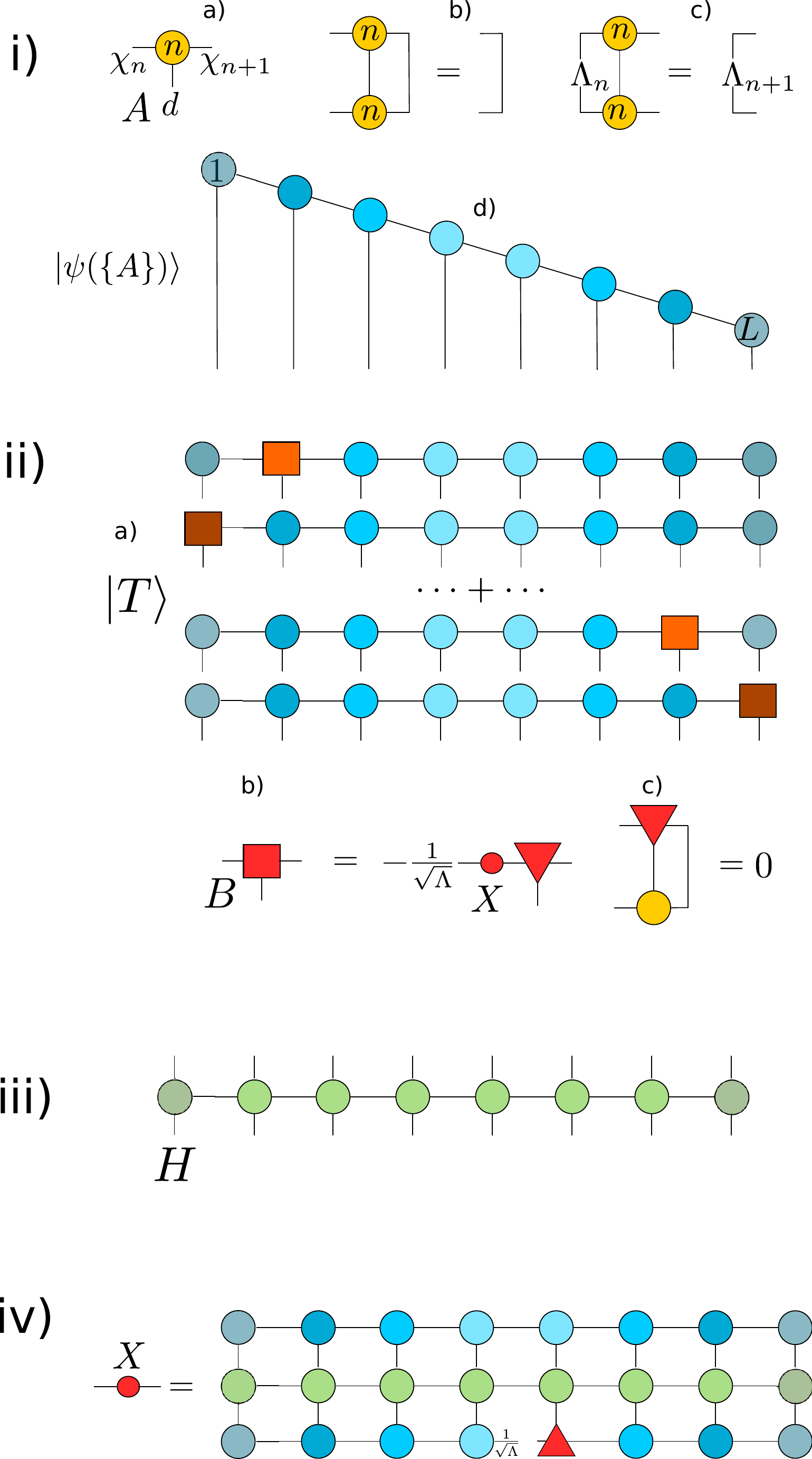}
\caption{\textbf{The TDVP for finite chains with MPO}. Geometric objects represent tensors, and lines attached to them represent their indexes that when connected are meant to be contracted. Lines interrupted by capital Greek letters represent diagonal matrices. \textbf{ i)} $A$ set of rank three tensor $A^{\alpha, \beta, i }_n$  where $n=1\cdots L$, $\alpha=1\cdots \chi_n, \ \beta=1, \cdots \chi_{n+1}, d$ (a), are required to  obey the gauge fixing conditions, so that they are isometric (b), and in the Schmidt basis, where the reduced density matrix $\Lambda$ is diagonal (c). In this way they are  used to construct a state $\ket{\psi(\{A\})}$ that has a well defined coarse-graining interpretation (d). \textbf{ ii)} A generic tangent vector $\ket{T}$ (a) is constructed from elementary rank three tensors $B_n$, $n=1\cdots L$ (b). The requirement of its orthogonality to the original state $\braket{\psi({A})|T}=0$ can be obtained by building all $B_n$ from a projector $V_n$, (triangle in the figure), a matrix of free coefficients $X_n$ and (for convenience) the inverse of the square root of the reduced density matrix $\Lambda_n$. The projector (c) is the one that ensures the orthogonality, by defining the $B_n$ in the orthogonal complement spanned by the respective $A$.  \textbf{ iii)} The Hamiltonian can be encoded in a MPO following the recipes in the literature \cite{mcculloch_infinite_2008, crosswhite_applying_2008, frowis_tensor_2010,pirvu_matrix_2010}. \textbf{ iv)} The TDVP amounts to solving the Schr\"odinger equation \ref{eq:Schr} projected on to the tangent plane, and this implies that the $X_n$ defining the $B_n$ variations used to update the $A_n$  at a specific step of the optimization are defined as in the drawing. \label{fig:tdvp}    }

\end{figure}

In oder to extract the best possible MPS description of the ground state of a given Hamiltonian there are several algorithms. Each of them  present some advantages, i.e. the original tebd is very used given is  simplicity \cite{vidal_efficient_2004} while variational methods based on the energy minimization are often preferred since they are faster \cite{mcculloch_infinite_2008}.
Most of the MPS based algorithms (including the original DMRG proposals \cite{white_density_1992}) rely on the fact that the MPS bond dimension $\chi$ grows during the computation (typically from $\chi$ to $ d \chi$ where $d$ is the dimension of the local Hilbert space), and is then reduced again to $\chi$ by keeping only the biggest singular values of a specific bi-partition of the system \cite{vidal_efficient_2004}. Recently new strategies have been developed based on the geometric notion of the MPS tangent plane \cite{haegeman_time-dependent_2011} that allow to optimize the MPS by solving a differential equation, without the need of extending its bond dimension. Here we describe how to implement this strategy for finite chains (see also \cite{verschelde_variational_2011}) and for Hamiltonians encoded in MPOs . We define a state $\ket{\psi\{A\}}$ generated as a MPS from a set $A_i, i=1\cdots L$ of rank three tensors in the standard way. The first important part of the algorithm consists in choosing a gauge for the MPS matrices. We work in the isometric gauge defined in Fig. \ref{fig:tdvp} i). In this way it is possible to give to the MPS an RG interpretation. The tensor $A_n$ at site $n$ basically coarse grain the block on the right of $n$ with the site $n$ and project it to a subspace of the Hilbert space. It has dimensions $\chi_n, \chi_{n+1}, d$ and  projects the tensor product Hilbert space built from ${\cal C}^{\chi_{n+1}}\otimes d$ into the Hilbert space ${\cal C}^{\chi_{n}}$ relevant for the description of the state.
 At each site of the chain one can define $\chi_{n+1}*d-\chi_{n}$ MPS tangent vectors, see \cite{haegeman_time-dependent_2011}. The requirement that those vectors are orthogonal to the original MPS vector is imposed by defining them through the projection onto the part of the  ${\cal C}^{\chi_{n+1}}\otimes d$ discarded for the description of the original  state, that has indeed the correct dimension  $\chi_{n+1}*d-\chi_{n}$. From a practical point of view, we can think of the tangent vector as a linear superposition of $L$ MPS states where for each of them one of the original tensor $A_n$  has been replaced by a new tensor $B_n$ as sketched in Fig.\ref{fig:tdvp} ii) . In order to ensure the orthogonality, the $B_n$ tensor are defined as the contraction of auxiliary tensors, (for normalization convenience ) the inverse square root of the reduced density matrix, times a matrix of free coefficients of dimension   $\chi{n}, \chi_{n+1}*d-\chi_{n}$ called  $X_n$, and a fixed projector $V_n$, that is the ultimate responsible of the orthogonality (see Fig.\ref{fig:tdvp} ii) c) ).
In  order to deal with the LR interactions, we encode the Hamiltonian in an MPO. Unfortunately MPOs cannot encode exactly polynomially decaying interactions, so that one needs to approximate the desired power law with a series of exponentials (for details see \cite{mcculloch_density-matrix_2007,pirvu_matrix_2010,crosswhite_applying_2008, frowis_tensor_2010,mcculloch_infinite_2008}). The graphical representation of the MPO encoding the Hamiltonian is given in Fig. \ref{fig:tdvp} iii).
If we want to obtain the ground state of a given Hamiltonian,  we can now start with a random MPS and solve the Schr\"oedinger equation in imaginary time for very long times. In formula we would like to solve for $ \ket{\psi(\{A(t)\})}$ the long time 
\begin{equation}
 \partial_t \ket{\psi(\{A(t)\})} = - H \ket{\psi(\{A(t)\})}. \label{eq:Schr}
\end{equation}
A possible way to do it is to project the equation  onto the  tangent plane defined as the collection of     $\chi_{n+1}*d-\chi_{n}, n=1\cdots L$ vectors, \footnote{Whenever this expression is zero, it just means that the tangent space is zero dimensional at that specific point}. In formula we would like to find the tangent vector $\ket{T}$ that minimizes the distance from  $H \ket{\psi(\{A(t)\})}$,
\begin{equation}
 \ket{T^*}, : min_{\ket{T}} || \ket{T} - H \ket{\psi(\{A(t)\})}||^2.
\end{equation}
The optimal $\ket{T*}$ is built from a collection of $\{T_n^*, n=1\cdots L\}$ that are used to update the $A_n$,

\begin{equation}
 A_n(t+dt) = A_n(t) - dt  T_n^*.
\end{equation}

At the end of each step the MPS state should be brought back to the original isometric gauge, and the procedure is iterated up to convergence.
From the computational point of view the definition of the tangent vectors as being orthogonal to the MPS state involves several simplifications. In particular it implies that one can build the $X_n$ matrices directly form the Hamiltonian and the $A$ tensors as written explicitly in Fig . \ref{fig:tdvp} iv).

\end{document}